\documentclass[aps,prx,twocolumn,superscriptaddress,longbibliography]{revtex4-1}
\usepackage{amssymb}
\usepackage{amsmath}
\usepackage{graphicx}
\usepackage{xcolor, ulem}
\usepackage[version=3]{mhchem}
\usepackage{footnote}
\usepackage{bm}
\usepackage{dsfont}
\usepackage{xr-hyper}
\usepackage{hyperref}
\usepackage{cleveref}
\crefname{equation}{Eq.}{Eqs.}
\Crefname{equation}{Equation}{Equations}
\crefname{figure}{Fig.}{Figs.}
\Crefname{figure}{Figure}{Figures}
\crefname{section}{Sect.}{Sects.}
\Crefname{section}{Section}{Sections}
\crefname{table}{Table}{Tables}
\crefname{appsec}{Appendix}{Appendices}
\graphicspath{{Figures/}}
\definecolor{adpcolor}{rgb}{0.36,0.54,0.66}

\definecolor{kvkcolor}{rgb}{1,0.5,0.05}

\usepackage[cal=boondox]{mathalfa}

\begin{document}

\textheight = 59\baselineskip

\title{Bifluxon: Fluxon-Parity-Protected Superconducting Qubit}

\author{Konstantin Kalashnikov}
 \email{kostya.kalashnikov@rutgers.edu}
\affiliation{Department of Physics and Astronomy, Rutgers University, Piscataway, NJ}

\author{Wen Ting Hsieh}
\affiliation{Engineering Department, University of Massachusetts Boston, Boston, MA}

\author{Wenyuan Zhang}
\author{Wen-Sen Lu}
\author{\\Plamen Kamenov}
\affiliation{Department of Physics and Astronomy, Rutgers University, Piscataway, NJ}

\author{Agustin Di Paolo}

\affiliation{Institut Quantique and D\'epartement de Physique, Universit\'e de Sherbrooke, Sherbrooke J1K 2R1 QC, Canada
}
\author{Alexandre Blais}
\affiliation{Institut Quantique and D\'epartement de Physique, Universit\'e de Sherbrooke, Sherbrooke J1K 2R1 QC, Canada
}
\affiliation{Canadian Institute for Advanced Research, Toronto, ON, Canada}
\author{Michael E. Gershenson}
\affiliation{Department of Physics and Astronomy, Rutgers University, Piscataway, NJ}

\author{Matthew Bell}
 \email{Matthew.Bell@umb.edu}
\affiliation{Engineering Department, University of Massachusetts Boston, Boston, MA}
\date{\today}

\begin{abstract}

We have developed and characterized a symmetry-protected superconducting qubit that offers simultaneous exponential suppression of energy decay from charge and flux noise, and dephasing from flux noise.
The qubit consists of a Cooper-pair box (CPB) shunted by a superinductor, thus forming a superconducting loop. Provided the offset charge on the CPB island is an odd number of electrons, the qubit potential corresponds to that of a $\cos{\left(\phi/2\right)}$ Josephson element, preserving the parity of fluxons in the loop via Aharonov-Casher interference. In this regime, the logical-state wavefunctions reside in disjoint regions of phase space, thereby ensuring the protection against energy decay. By switching the protection on, we observed a ten-fold increase of the decay time, reaching up to $100\,\mu$s. 
Though the qubit is sensitive to charge noise, the sensitivity is much reduced in comparison with the charge qubit, and the charge-noise-induced dephasing time of the current device exceeds $1\,\mu$s.
Implementation of the full dephasing protection can be achieved in the next-generation devices by combining several $\cos{\left(\phi/2\right)}$ Josephson elements in a small array.

\end{abstract}

\maketitle

\section{Introduction}

 Superconducting qubits have emerged as one of the most promising platforms for quantum computing \cite{kjaergaard2019superconducting}. Over the past two decades, the coherence of these qubits has been improved by five orders of magnitude \cite{devoret2013superconducting}. Even with this spectacular progress, implementation of error correction codes remains very challenging \cite{preskill2018quantum}. Further improvement in coherence will require the development of new approaches for mitigating harmful effects due to uncontrollable microscopic degrees of freedom, such as two-level systems (TLS) in the qubit environment \cite{muller2017towards}. This route is provided by the improvement of materials involved in fabrication of superconducting qubits, which can lead to the reduction of the TLS density. A complementary approach, which we consider below, is based on the reduction of the qubit-TLS coupling by qubit design.

Qubit coherence is characterized by the energy relaxation (decay) time $T_1$ and the dephasing time $T_\varphi$. The decay rate $\Gamma_1\equiv1/T_1$ due to coupling to a fluctuating quantity $\lambda$ is proportional to the transition amplitude $|\langle g| {H_\lambda}|e \rangle|^2$, where ${H_\lambda}$ is the coupling Hamiltonian and $\{|g\rangle,|e\rangle\}$ are the qubit's logical states. Since the external noise couples to local operators, decreasing of the overlap of $|g\rangle$ and $|e\rangle$ wavefunctions can significantly reduce $\Gamma_1$. This strategy is exploited by several qubit designs in which localization of the logical-state wavefunctions occurs within distinct and well-separated minima of the qubit potential, such as the ``heavy fluxonium'' qubit \cite{earnest2018realization, lin2018demonstration}.

\begin{figure}[t!]
\includegraphics[width=\columnwidth]{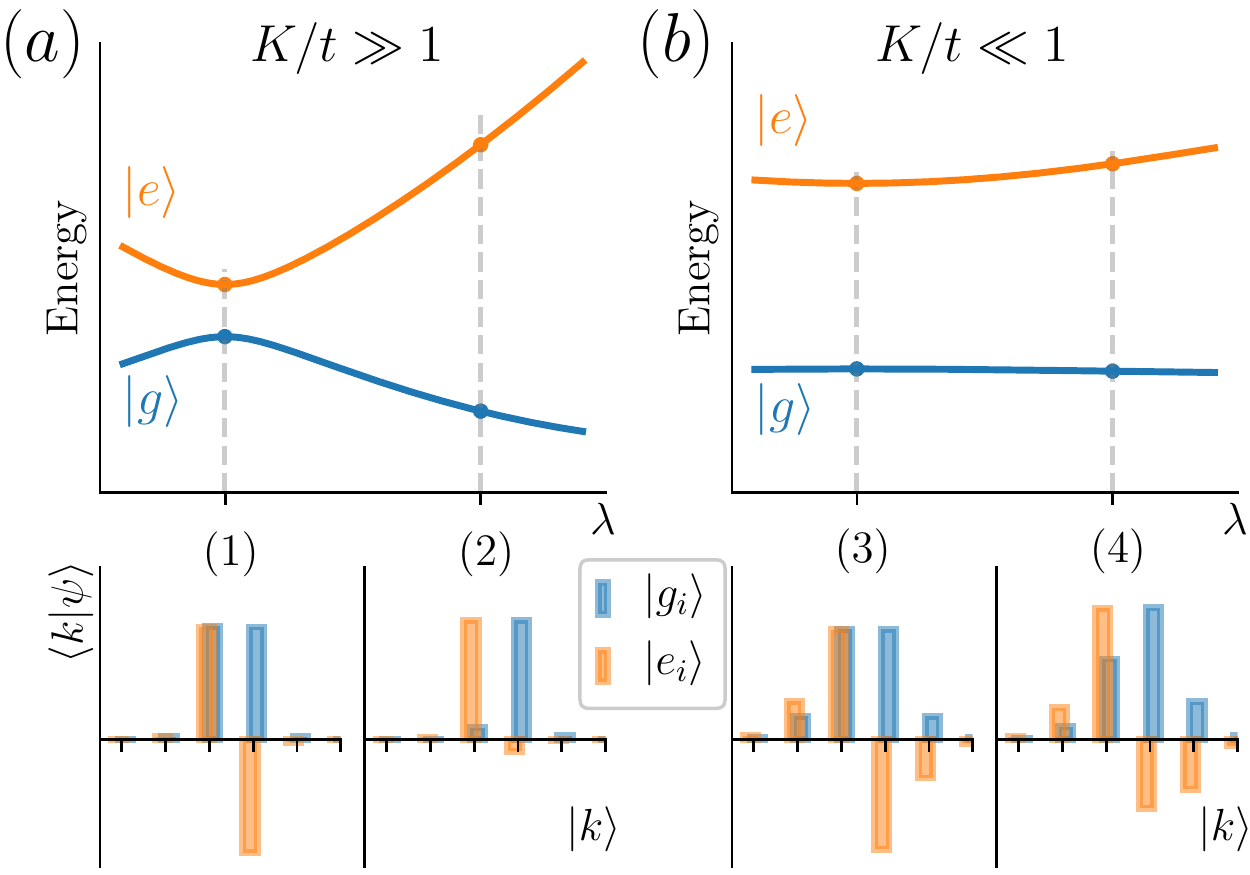}
\caption{\label{fig:cpb_wfs} The trade-off between the decay and dephasing protection in superconducting qubits with a single charge or flux degree of freedom. The band structure (top panels) and wavefunctions (bottom panels) of a particle in quasiperiodic potentials: (a) the free-particle regime and (b) the tight-binding regime. The wavefunction overlap and the energy sensitivity $ \partial E_{eg}^{(i)}/ \partial \lambda$ do not simultaneously vanish for any point $(i)$. Flux (charge) qubits correspond to the case of the control parameter $\lambda = \Phi_{\mathrm{ext}}$ $(q_g)$, kinetic energy $K = E_L$ $(E_C)$, tunneling energy $t = E_{sps}$ $(E_J)$, and  $|k\rangle$ as a fluxon (charge) basis.}
\end{figure}

On the other hand, a small dephasing rate $\Gamma_\varphi\equiv1/T_\varphi$ requires the qubit  transition frequency $\omega_{ge}$ to be insensitive to fluctuations of $\lambda$. The first-order decoupling of a qubit from noise has been achieved at the so-called ``sweet spot'' $\lambda_0$, where $\partial\omega_{ge}/\partial\lambda|_{\lambda_0}=0$ \cite{vion2002manipulating}. However, the coherence times achieved with this approach are insufficient for the implementation of the error correction codes, even if  the drifts of the qubit operating point are eliminated over the timescale of operations. To remedy this, a ``sweet-spot-everywhere'' approach has been realized in the transmon qubit \cite{houck2009life, koch2007charge}: an exponentially strong suppression of the qubit sensitivity to noise has been achieved by delocalization of the qubit wavefunctions in charge space.

It is, however, worth noticing that the two approaches of $T_1$ and $T_{\varphi}$ protection by qubit design come into conflict in the case of devices with a single degree of freedom in the qubit Hamiltonian (which we will refer to as 1D qubits). For instance, at the dephasing sweep spot of the ``heavy fluxonium'' \cite{earnest2018realization, lin2018demonstration} wavefunctions become delocalized due to its hybridization, which limits decay time [\cref{fig:cpb_wfs}(a), $i=1$], whereas $T_1$ protection can be realized only at the slope of dispersion curve where  $T_{\varphi}$ is small. [\cref{fig:cpb_wfs}(a), $i=2$]. In turn, the charge insensitivity of the transmon qubit is accompanied with strong dipole matrix elements that limit $T_1$ [\cref{fig:cpb_wfs}(b), $i=3,4$]. Additionally, the flatness of the transmon-qubit bands results in a strong reduction of the spectrum anharmonicity, potentially leading to a leakage of information outside of the computational subspace \cite{brink2018device}.

These examples suggest that a qubit Hamiltonian with full noise protection against relaxation and dephasing, i.e. exponentially large $T_1$ and $T_{\varphi}$, cannot be implemented in a single-mode superconducting quantum device. This conflict, however, can be reconciled by the so-called ``few-body'' qubits \cite{smith2019superconducting}, that incorporate more than one degree of freedom in the qubit Hamiltonian (the dimentionality $D > 1$) \cite{kitaev2006protected,  bell2014protected, dempster2014understanding, kou2017fluxonium}.

An example of simultaneous decay and dephasing protection in circuits with $D>1$ is given by the $0$-$\pi$ qubit \cite{brooks2013protected}. Its $D=2$ Hamiltonian combines one ``light'' $\phi$ and one ``heavy'' $\theta$ variable. The logical wavefunctions are delocalized along the $\phi$ direction, while being localized in two disconnected potential wells labeled by $\theta =[0,\pi]$. These properties lead to exponentially reduced sensitivity to flux-noise fluctuations, i.e. negligible dephasing, and exponentially small matrix elements, i.e. long decay time \cite{groszkowski2018coherence}. Noise protection in this device is hard-wired by circuit design, making the qubit robust against external perturbations. Fabricating such a circuit, however, entails several serious challenges, among which are very strict requirements on the parameters of all circuit elements and symmetry constraints. Moreover, since the built-in protection permanently decouples the qubit from the environment, new approaches to state preparation, qubit manipulation, and readout are required \cite{di2019control}. 

Another concept of qubit protection exploits symmetries of Hamiltonians with $D>1$ \cite{ douccot2012physical}, an example being the qubit based on Josephson rhombi arrays \cite{douccot2002pairing}, experimentally realized in Ref. \cite{bell2014protected}. In a single rhombus threaded by half of the magnetic flux quantum, the transport of individual Cooper pairs (CP) is suppressed due to destructive Aharonov-Bohm interference, such that the rhombi chain supports correlated transport of CP pairs [i.e., acts as a $\cos(2\phi)$ Josephson element]. The dephasing time of the qubit can be enhanced by delocalization of wavefunctions over the states with the same CP parity, which does not compromise $T_1$. Importantly, this qubit design enables on-demand tuning the qubit coupling to the environment (including the read-out) on and offs, which facilitates qubit manipulations. This also provides a route to fault-tolerant gates immune to noises in the control lines \cite{klots2019set}. An improved version of the rhombus qubit can be built by parallel connection of several rhombi chains \cite{gladchenko2009superconducting}.

\begin{figure}[t]
\includegraphics[width=\columnwidth]{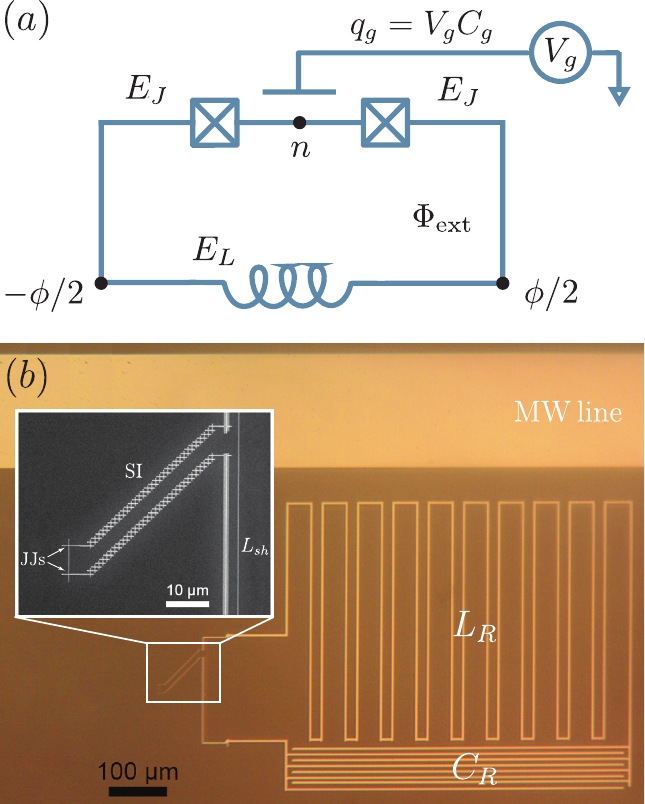}
\caption{\label{fig:bifluxon_simple} (a) Simplified circuit scheme of the bifluxon qubit, described by \cref{eq:full H}. Charging energies of the superinductor and CPB are $E_{CL}$ and $E_{C}$, respectively. The qubit is controlled by the CPB charge $q_g$ and the magnetic flux $\Phi_{\mathrm{ext}}$. (b) Optical image of the bifluxon qubit, readout resonator, and the microwave (MW) transmission line. The inset shows the SEM image of its central part: two JJs form CPB island, the long array of larger JJs acts as a superinductor.   }
\end{figure}

Here we focus on the implementation of a complementary circuit preserving the parity of fluxons in a superconducting loop, which consists of a split Cooper-pair box (CPB) and a superinductor (SI), and is depicted in \cref{fig:bifluxon_simple}(a). The probability of single fluxon tunneling in and out of the loop can be tuned by the CPB charge $q_g$ of the CPB island (hereafter we refer to CPB charge modulo 2 due to periodicity).
At $q_g = 1e$ (where $e$ is the electron charge) Aharonov-Casher interference results in a $4\pi$-periodic potential [i.e. $\cos(\phi/2)$ Josephson element], which preserves the fluxon parity in the loop \cite{friedman2002aharonov, bell2016spectroscopic, de2018charge}.
In the case of perfectly symmetric CPB junctions, the two degenerate logical states with different fluxon-number parity reside in disjoint regions of the Hilbert space, forbidding qubit decay. It is moreover possible to delocalize the wavefunction within each parity state via double fluxon tunneling in order to provide protection against pure dephasing by flux noise. Below we refer to such an element as a ``Bifluxon'' qubit.

In this paper, we have designed and characterized a prototype of the bifluxon qubit and demonstrated the decay protection by setting the CPB charge to the value of $1e$. By turning protection on, we observe a ten-fold increase of the decay time, up to 100 $\mu$s. We also report the measurement of the qubit phase-coherence time $T_\varphi$, exceeding $ 1\,\mu$s. 

The paper is organized as follows. In \cref{smain:Theory} we elaborate on the coherence properties of the bifluxon qubit by analyzing the symmetries of the logical wavefunctions and the resulting selection rules, as well as possible ways to realize dephasing protection. In \cref{smain:Experiment} we present experimental implementation of the bifluxon qubit and discuss coherence-time measurement protocols.  In \cref{smain:Discussion} we analyse the coherence limitations of the bifluxon qubit, and  discuss a number of possibilities for further coherence-time improvements. 

\section{Theory}
\label{smain:Theory}

This section outlines the theory of the bifluxon qubit and the origin of its noise protection. We assume for simplicity that the Josephson junctions (JJs) forming the CPB are identical with Josephson energy $E_J$ and charging energy $E_C$. The charging and inductive energies of the superinductance $L$ are denoted by $E_{CL}$ and $E_L=(\Phi_0/2\pi)^2 / L$, respectively, where $\Phi_0 = h/2e$ is the  quantum of magnetic flux. 

The behavior of the system is determined by two controllable parameters: the offset charge $q_g$ of the CPB island and the external flux $\Phi_{\mathrm{ext}}$ through the device's loop. In this section we will use the dimensionless quantities $\varphi_{\mathrm{ext}} = 2 \pi \Phi_{\mathrm{ext}}/\Phi_0$ and $n_g =q_g/2e$, and restore  the dimensionful values in the experimental section. The circuit Hamiltonian has three degrees of freedom (see \cref{sapp:Derivation of the circuit Hamiltonian}): the superconducting phase of the CPB island, $\varphi$, and the sum and difference of the phases at the ends of the superinductor, respectively denoted by $\phi_+$ and $\phi$. For simplicity, we assume that the high-frequency circuit mode $\phi_+$ is not excited. Under this approximation, the qubit Hamiltonian is two-dimensional. In the charge basis for the CPB degree of freedom, the circuit Hamiltonian can be written as 
\begin{equation}
\begin{split}
H = &  \sum_{n} \Big[4E_C(n-n_g)^2 |n\rangle \langle n| - E_J \cos (\phi/2)(\sigma_n^+ + \sigma_n^-)\Big] \\
& - 4E_{CL} \partial_\phi^2 + \frac{E_L}{2}(\phi-\varphi_{\mathrm{ext}})^2,
\end{split}
\label{eq:full H}
\end{equation}
where $n$ represents the number of Cooper pairs in the CPB island, and we have defined  $\sigma_n^+=|n+1\rangle\langle n|$ and  $\sigma_n^- = (\sigma_n^+)^{\dagger}$.

\begin{figure*} [pt]
\includegraphics[width=\textwidth]{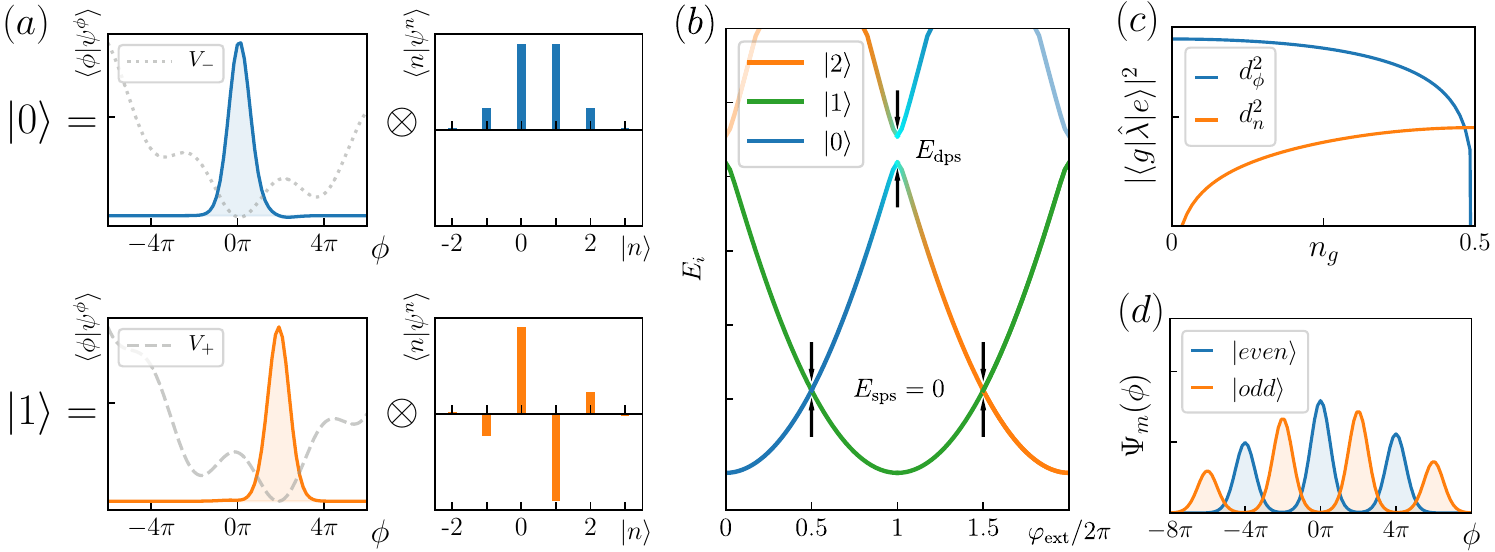}
\caption{\label{fig:theory-1} 
(a) The ground and first excited states of the bifluxon qubit shown as products of the fluxon wavefunctions in the $V_{\pm}$ potentials and CPB state, for $\varphi_{\mathrm{ext}} = \pi, \, n_g = 0.5$. The parity of the $\cos{(\phi/2)}$ term is controlled by the CPB state $|\pm^n \rangle$. (b) Bifluxon energy bands as a function of $\varphi_{\mathrm{ext}}$ at $n_g =0.5$. Color gradient represents the hybridization of the states with different fluxon numbers. Note the crossing of the parabolas at half-integer $\varphi_{\mathrm{ext}}/2\pi$ due to $E_{\mathrm{sps}} = 0 $ and the avoided crossing between the next-to-neighbor parabolas $E_{\mathrm{dps}}>0$. (c) Dependence of the flux and charge matrix elements on the CPB charge. The decay protection is realized at $n_g = 0.5$, where $d_{\phi}$ is zero and $d_{n}$ is significantly suppressed. (d) Delocalization of  the even (odd) fluxon states in the regime $E_{\mathrm{dps}} \gg 2\pi^2 E_L$ leads to suppression of dephasing due to the flux noise. }
\end{figure*}

To illustrate the working principles of the bifluxon qubit, here we examine the limiting case $E_C \gg E_J$, although full numerical diagonalization is used to analyze the data from devices with $E_J \gtrsim E_C$ below. Let us consider two cases for the offset charge $n_g$. If $n_g$ is set near an integer number $N$, the CPB degree of freedom can be thought as ``frozen'' close to the charge state that minimizes the kinetic-energy term in \cref{eq:full H}. In this case, the circuit Hamiltonian is reduced to a 1D fluxonium-like Hamiltonian with a renormalized Josephson energy $E_J^2/4E_C$ (see \cref{sapp:Integer charge on the island}). To operate the bifluxon qubit in the protected regime, the offset charge instead should be set close to half-integer, i.e. $n_g\approx  1/2$. With $E_C \gg E_J$, it is sufficient to consider only two nearly degenerate CPB states $|0\rangle$ and $|1\rangle$. Projecting the circuit Hamiltonian in this two-dimensional subspace, we find 
\begin{equation}
\begin{split}
H_{r} = & 4 E_C\left(\frac{1}{2}-n_g\right)\sigma^z - E_J \cos (\phi/2)\sigma^x \\
& - 4E_{CL} \partial_\phi^2 + \frac{E_L}{2}(\phi-\varphi_{\mathrm{ext}})^2,
\end{split}
\label{eq:reduced H}
\end{equation}
where $\sigma^z=|1\rangle\langle 1|-|0\rangle\langle 0|$ and $\sigma^x=\sigma^+ + \sigma^-$. \cref{eq:reduced H} is diagonal in the $\sigma^x$ basis for $n_g=1/2$. Therefore, the lowest-energy eigenstates can be factorized as $|\psi^n \rangle \otimes |\psi^{\phi} \rangle$, where the superscripts $n$ and $\phi$ denote the charge- and flux-like components of the wavefunctions, respectively. In particular, the charge-like component results in either the symmetric or anti-symmetric combinations $ |\pm^n \rangle = \left(|0 \rangle \pm| 1 \rangle \right)/\sqrt{2}$. The flux-like component is an eigenstate of the one-dimensional Hamiltonian $H_{\pm}= - 4E_{CL} \partial_\phi^2 + V_{\pm}$, with a potential energy that depends on the charge state
\begin{equation}
V_{\pm} = \mp E_J\cos(\phi/2) + \frac{E_L}{2}(\phi-\varphi_{\mathrm{ext}})^2.
\label{eq:potetial}
\end{equation}
The local minima of the fluxonium-like potential $V_+$ ($V_-$) are positioned near $\phi_{m} = 2\pi m$, where $m$ is an even (odd) integer. An harmonic-oscillator wavefunction of the form $\psi_m(\phi) \sim \exp(-\sqrt{{E_J}/{E_{CL}}}\,(\phi-\phi_m)^2/4)$, localized at $m$-th minimum, can be associated with a fluxon excitation $|m\rangle$. Using the fluxon representation,  the eigenstates of \cref{eq:reduced H} can be expressed as $|m\rangle = \{ | 2k \rangle \}\cup\{ | 2k+1 \rangle  \}$, where $|2k\rangle = | +^n, \psi^{\phi}_{2k} \rangle$ and $|2k+1 \rangle = | -^n, \psi^{\phi}_{2k+1} \rangle $ have an even and odd number of fluxons in the loop, respectively [see \cref{fig:theory-1}(a)].

\Cref{fig:theory-1}(b) presents the spectrum of the qubit for $n_g=1/2$ as a function of $\varphi_{\mathrm{ext}}$. Since the single phase-slip (SPS) processes connecting $|m\rangle\leftrightarrow |m+1\rangle$ are forbidden due to the symmetry of the wavefunctions
\begin{equation}
E_{\mathrm{sps}} = \langle m| H_r |m+1 \rangle \propto  \langle +^n  | -^n  \rangle = 0,
\label{eq:sps}
\end{equation}
the two neighboring parabolas cross at half-integer $\varphi_{\mathrm{ext}}/2 \pi$. This can be interpreted as a fluxon-parity conservation rule due to the Aharonov-Casher effect \cite{friedman2002aharonov}, which has been experimentally observed in Refs. \cite{bell2016spectroscopic, de2018charge, pop2010measurement}. Therefore, at a half-integer $n_g$, the considered system resembles a fluxonium qubit made up of a $4\pi$-periodic Josephson element, justifying the name ``bifluxon''. 

Double phase-slip (DPS) processes mix fluxon states with the same parity ($m$ and $m+2$), opening energy gaps in the spectrum. The DPS amplitude is given by 
\begin{equation}
\begin{split}
E_{\mathrm{dps}} & =  \langle m| H_r |m+2 \rangle = \hbar \omega_p \langle \psi^{\phi}_{m}  | \psi^{\phi}_{m+2}  \rangle \\
& \approx \hbar \omega_p \exp(-\pi^2 \beta),
\end{split}
\label{eq:dps}
\end{equation}
where $\omega_p = \sqrt{8 E_J E_{CL}}$ is a plasma frequency for the $V_{\pm}$ potentials and $\beta = \sqrt{2E_J/E_{CL}}$.

The symmetry of states with distinct fluxon parity makes the qubit immune to energy decay due to both flux and charge noises. Indeed, the phase dipole-moment matrix element is identically zero
\begin{equation}
d_\phi \sim \langle m |\hat{\phi}| m+1 \rangle \propto \langle+^n |-^n \rangle = 0,
\label{eq:phase dipole}
\end{equation}
while, provided $E_J\gg E_{CL}$, the matrix element of the charge-noise operator is exponentially suppressed in comparison with the charge qubit \cite{astafiev2004quantum} 
\begin{equation}
d_n \sim \langle m |\sigma^z | m + 1 \rangle = \langle \psi^{\phi}_{m}  | \psi^{\phi}_{m+1} \rangle = \exp(-\pi^2\beta/4).
\label{eq:charge dipole}
\end{equation}
\Cref{fig:theory-1}(c) shows the charge and phase dipole-moment matrix elements obtained by numerical diagonalization of the full Hamiltonian \cref{eq:full H}. The weak sensitivity to charge noise is comparable to the flux sensitivity of a heavy fluxonium \cite{lin2018demonstration, earnest2018realization}, and can be suppressed by stronger localization of the single-well excitations within the $V_{\pm}$ potential minima by increasing the $E_J/E_{CL}$ ratio. 

The decay protection due to symmetries of the bifluxon-circuit wavefunctions can also be understood in the following way. Consider a logical qubit made of two faulty qubits labelled $\sigma$ and $\tau$, with the two lowest-energy states $| g \rangle =  |\uparrow^\sigma\downarrow^\tau \rangle$ and $ | e \rangle = | \downarrow^ \sigma\uparrow^\tau \rangle$ separated from the others by a sizeable energy gap $\Delta E$. Since uncorrelated fluctuations of $\sigma$ and $\tau$ cannot induce $g\leftrightarrow e$ transition and the leakage out of computation space is penalized by $\Delta E$, the qubit is protected against local noise in the $\sigma$ and $\tau$ subsystems. Accordingly, the bifluxon qubit is protected against decay due to uncorrelated charge and flux noises.
 
In addition to the decay protection, the bifluxon qubit can also be robust to flux-noise dephasing. Indeed, similarly to the case of the fluxonium qubit, the flux dispersion of the qubit can be reduced by increasing the superinductance value. This enables wider delocalization of the qubit wavefunctions in disjoint subspaces with different fluxon parities, as shown in \cref{fig:theory-1}(d). Quantitatively, the wavefunctions spread out over $G  \approx \sqrt{E_{\mathrm{dps}}/
E_L}$ potential wells. The flux dispersion is then suppressed by a factor of $\exp(-G)$ for $G \gg 1$. Therefore, the bifluxon qubit becomes exponentially insensitive to flux-noise dephasing under the condition $E_{\mathrm{dps}} \gg 2\pi^2 E_L$. Although this requirement is a challenge for the current fabrication capabilities, the implementation of ultrahigh-impedance superinductors has already been demonstrated \cite{bell2012quantum,pechenezhskiy2019quantum}.

Finally, it should be noted that, since the bifluxon is inherently a charge-sensitive device, a single qubit does not offer a protection against the charge-noise-induced dephasing. As we will discuss in \cref{smain:Discussion}, small array of such elements can in principle provide a polynomial increase of the dephasing time and help to overcome this limitation.

\section{Experiment}
\label{smain:Experiment}

\begin{figure}[th!]
\includegraphics[width=\columnwidth]{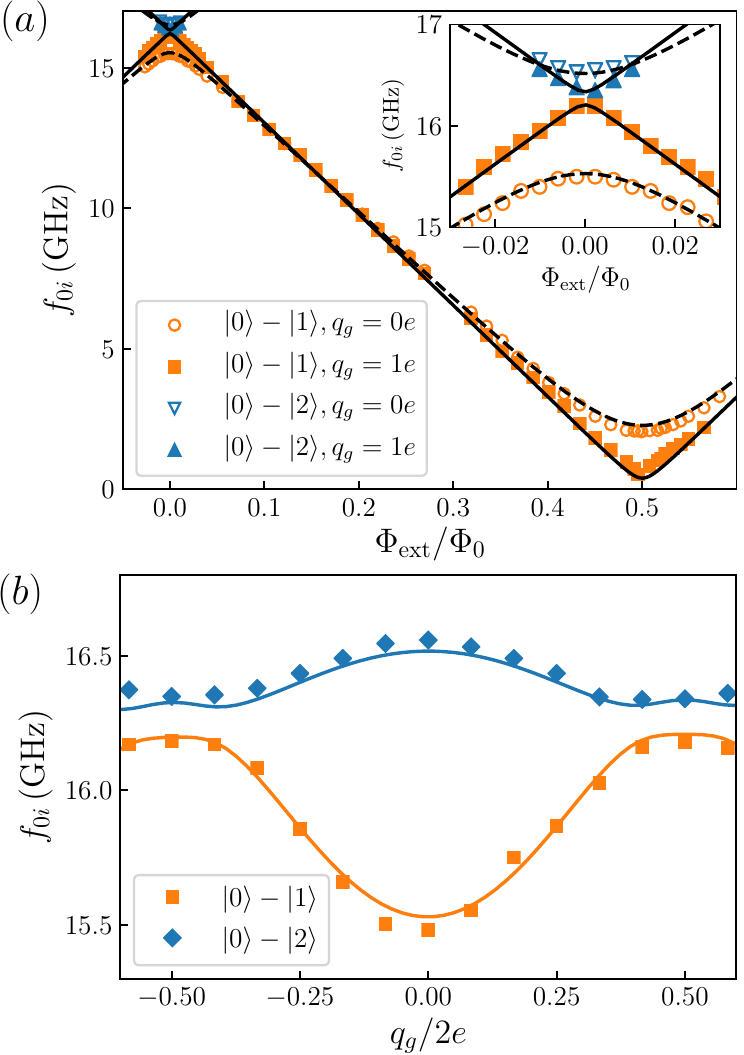}
\caption{\label{fig:spectra} Spectra of the bifluxon qubit: experimental data for the $|0\rangle-|1\rangle$  and $|0\rangle-|2\rangle$ transitions (symbols) and the result of exact diagonalization of the circuit Hamiltonian in \cref{eq:full H} (solid lines). (a) Flux dispersion of the transition frequencies $f_{0i}$ for two values of the CPB charge $q_g= 0, 1 \,e$. The inset is a zoom in of the qubit spectrum near $\Phi_{\mathrm{ext}} = 0$, displaying the avoided crossing that characterizes the rate of double phase slips $E_{dps}$. (b) Charge dispersion of the $f_{0i}$ transition frequency for $\Phi_{\mathrm{ext}} = 0$.  }
\end{figure}

In this work, the bifluxon qubit is realized as a split-junction CPB [a superconducting island flanked by two small nominally identical JJs with Josephson and charging energies $E_J$ and $E_{CJ}$, respectively; see \cref{fig:bifluxon_simple}(b)] shunted by a superinductor (SI), which is implemented as an array of $N_A = 122$ larger JJs with corresponding energies $E_{JA}$ and $E_{CA}$. The sizes of small ($0.11 \, \mu \mathrm{m} \times 0.16 \, \mu \mathrm{m}$) and large ($0.21 \, \mu \mathrm{m} \times 0.30 \,  \mu \mathrm{m}$) junctions are chosen in order to allow phase-slip events across the CPB junctions ($E_J/E_{CJ} \sim 1$), but suppress the phase slips in the array ($E_{JA}/E_{CA} \gg 1$). As long as the inductive energy of the SI chain $E_L=E_{JA}/N_A$ is much smaller than $E_J$, the phase across the SI is close to integer number of $2\pi$. The stray capacitance of the superconducting islands to the ground in combination with the junction capacitances results in charging energies $E_{C}$ and $E_{CL}$ of the CPB and the SI, respectively (see \cref{sapp:Derivation of the circuit Hamiltonian} for details). The self-resonant mode of the SI with the frequency $\sim \sqrt{E_L E_{CL}}/h$ should remain well above the qubit transition frequency (usually $\sim \mathrm{few}\, \mathrm{GHz}$) in order to avoid qubit coupling to this mode.

The bifluxon qubit is controlled by the magnetic flux in the loop $\Phi_{\mathrm{ext}}$ and the offset charge $q_g$, induced by applying the dc bias voltage to the coupling capacitor $C_g$ between the microstrip line and the CPB island.
In order to perform the dispersive measurements of the bifluxon qubit, the device is inductively coupled to a lumped-element readout resonator with capacitance $C_R = 120\, \mathrm{fF}$ and inductance $L_R = 4 \, \mathrm{nH}$. For the coupling, a portion of the bifluxon superconducting loop with kinetic inductance $L_{sh} = 0.4\, \mathrm{nH}$ is shared with the readout resonator. The qubit-resonator coupling constant for the device described in this paper is found to be $ g/2\pi= 52 \, \mathrm{MHz}$. 

 In the transmission measurements, the microwave signals travel along the microstrip line which is coupled to the readout resonators of up to 5 different bifluxon qubits measured in the same cooldown. The qubits could be individually addressed due to different resonant frequencies of the read-out resonators. The bifluxon qubit, readout resonator, and microstrip transmission lines are fabricated in a single multi-angle electron-beam deposition of aluminum through a lift-off mask (for fabrication and measurement details, see Refs. \cite{bell2014protected, bell2016spectroscopic}). 
 
 The pump tone $f_p$ induces the $|0\rangle-|i\rangle$ transitions at the resonance frequencies $f_{0i} = (E_{i} - E_0)/h$. The measurement tone $f_m$ probes the dispersive shift of the coupled read-out resonator. Although the dispersive measurements in the protected regime are complicated by significantly reduced qubit-readout coupling, the signal-to-noise ratio in the spectroscopic measurements was sufficiently high to identify the resonances even in the protected regime.
 The flux dependences of the resonance frequences $f_{01}$ and $f_{02}$ at $q_g=0 , \, 1e$ are shown in \cref{fig:spectra}(a).
The obtained spectra are in a good agreement with the results of diagonalization of the circuit Hamiltonian [\cref{eq:full H}, solid lines in \cref{fig:spectra}(a)], with the fitting parameters $E_J/h=27.2 \, \mathrm{GHz}$, $E_C/h=7.7 \, \mathrm{GHz}$, $E_L/h=0.94 \, \mathrm{GHz}$, $E_{CL}/h=10 \, \mathrm{GHz}$, and asymmetry between the CPB junctions ${\Delta E}_J/h=6 \, \mathrm{GHz}$.

The extracted values are consistent with the expected JJ parameters. The normal-state resistance of the CPB junctions extracted from $E_J$ using the Ambegaokar-Baratoff relation agrees within $20 \%$ with the resistance of test junctions fabricated on the same chip. Both CPB and SI charging energies agree well with the typical aluminum-based junction capacitance $50 \, \mathrm{fF/}\mu \mathrm{m}^2$ and specific capacitance of micron-size islands on silicon substrates $0.04 \, \mathrm{fF/} \mu \mathrm{m}$ \cite{pozar2009microwave}. 

We also observed an additional resonance at $13.9\, \mathrm{ GHz}$, whose position did not depend on $\Phi_{\mathrm{ext}}$ and $q_g$.  We attribute this resonance to the lowest-frequency mode of the superinductor, which corresponds to characteristic impedance of the SI $Z = 14\, \mathrm{k}\Omega$. 

\begin{figure}[t!]
\includegraphics[width=\columnwidth]{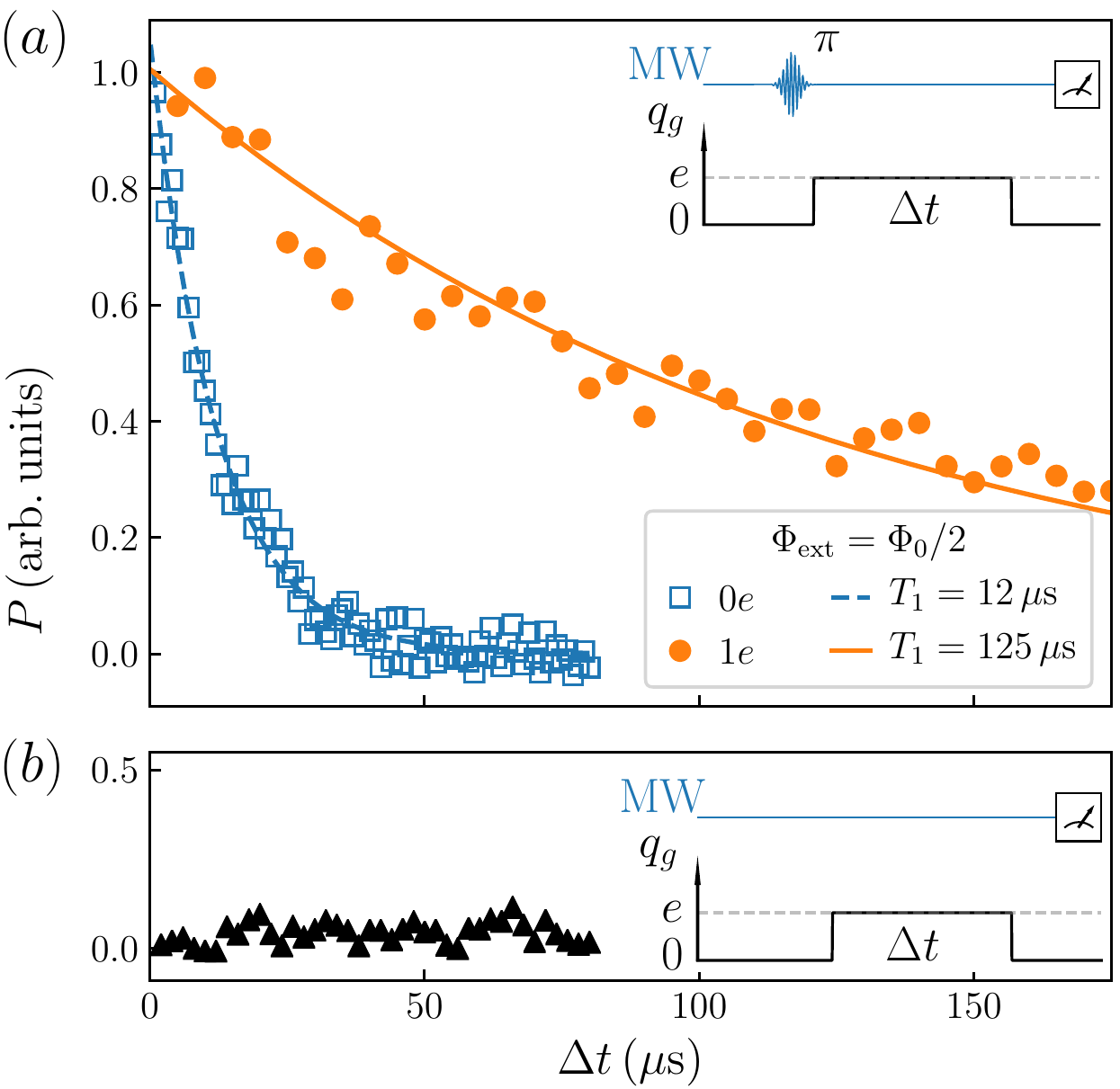}
\caption{\label{fig:decay} (a) Measurements of the bifluxon energy relaxation in the protected state (red circles) and unprotected state (blue squares). The sequence of pulses is shown in the inset. The exponential fits are shown by solid and dashed lines, respectively. (b) Demonstration of an absence of qubit excitation by the gate voltage pulses.}
\end{figure}

\begin{figure}[t!]
\includegraphics[width=\columnwidth]{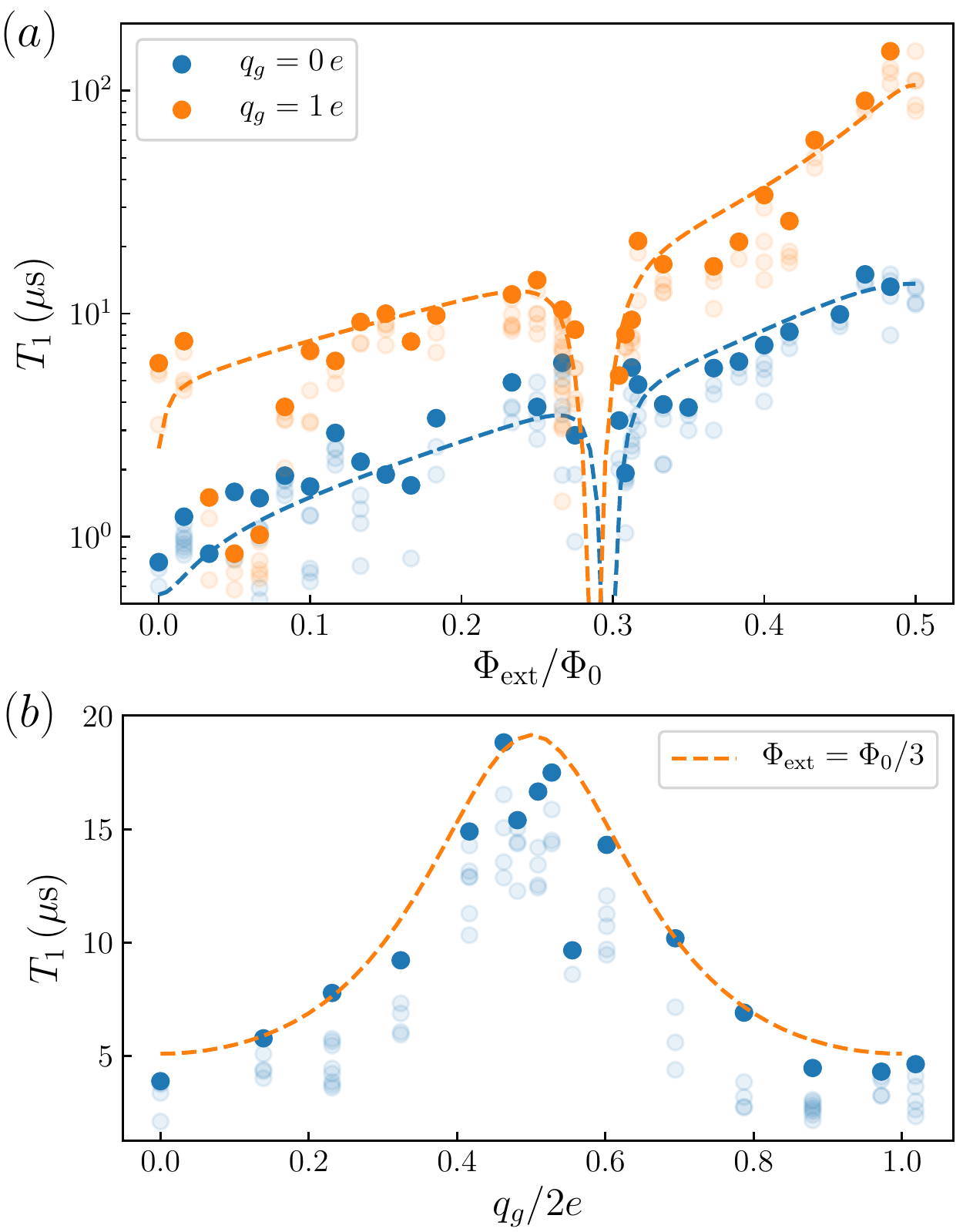}
\caption{\label{fig:T1_summary} Energy relaxation time $T_1$ as a function of the flux frustration $\Phi_{\mathrm{ext}}/\Phi_0$ (a) and the CPB charge $q_g$ (b). The pale dots represent all the measured data, the bright dots show the longest $T_1$ measured for a given operation point. The dashed lines correspond to fitting to the resistive noise theory (\cref{sapp:Coupling to the environment and decoherence}).}
\end{figure}

 In the time-domain experiments the signal-to-noise ratio, reduced by weak qubit-readout coupling, is too low to employ conventional pulse protocols (decay, Rabi oscillations and Ramsey fringes). For this reason we designed special pulse sequences for $T_1$ and $T_2$ measurements in the protected regime.
The pulse sequence used for probing the decay is show in  \cref{fig:decay}(a). Initially the qubit is prepared in the ground unprotected state ($q_g=0e$). A microwave $\pi$-pulse at the resonant frequency $f_{01}^{(0e)}$ excited the qubit, and then the protection is turned on by applying a pulse of the gate voltage $V_g$ corresponding to the offset charge $q_g = 1e$. We have used $V_g$ pulses with the rise/drop time $ \sim 30 \, \mathrm{ns}$, which is sufficiently long to ensure adiabatic evolution of the qubit between protected and unprotected states. After time $\Delta t$, the protection is removed by setting $q_g=0$ and the qubit state is measured. As a control experiment, we apply the gate voltage pulses alone, without a $\pi$-pulse; the absence of qubit excitation proved the adiabaticity of gate manipulations, see \cref{fig:decay}(b).

The main result of this paper - the dependence of $T_1$  on the qubit control parameters $\Phi_{\mathrm{ext}}$ and $q_g$ - is presented in \cref{fig:T1_summary}. Dashed lines represent fits to the model that takes into account resistive losses in the capacitevely coupled environment and readout resonator (Purcell effect). The details of the $T_1$ calculations are provided in \cref{sapp:Coupling to the environment and decoherence}. An increase of $T_1$ in the protected regime by an order of magnitude provides evidence for the qubit's dipole moment suppression. The longest decay time $>100 \, \mu \mathrm{s}$ is measured at full flux frustration $\Phi_{\mathrm{ext}} = \Phi_0/2$, which corresponds to a minimum qubit energy $f_{01}^{(1e)} = 0.4 \, \mathrm{GHz}$. The routs to further increase of $T_1$ are discussed in \cref{smain:Discussion}. 

\begin{figure}[t!]
\includegraphics[width=\columnwidth]{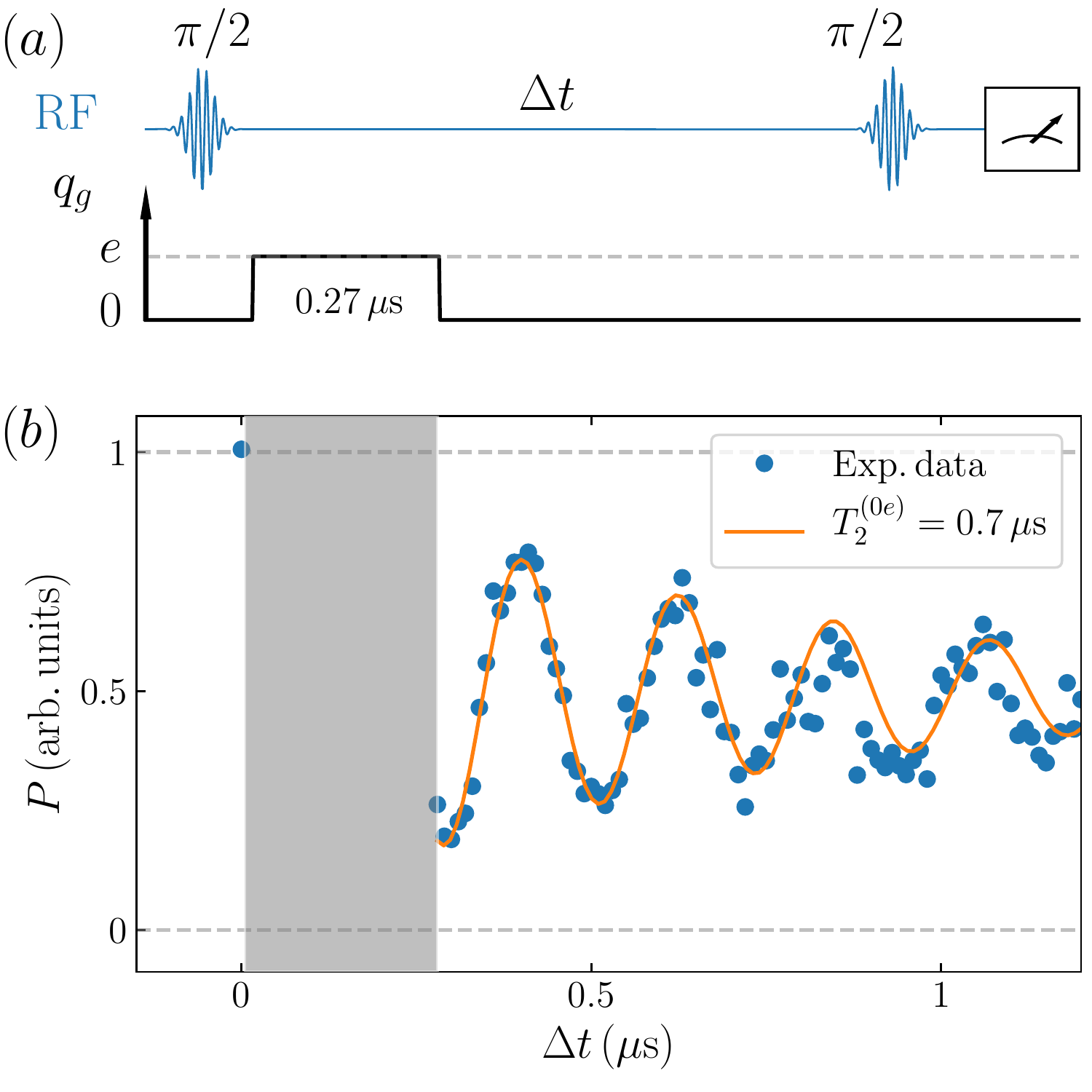}
\caption{\label{fig:T2} The Ramsey fringes measurement. (a) The pulse protocol for $T_2$ evaluation in the protected state. The  protection is turned on for a fixed time of $270\, \mathrm{ns}$; the time delay between two $\pi/2$ pulses is varied in order to record Ramsey fringes. (b) The experimental data (dots) and the damped-oscillation fitting (the solid line). Note that the value of $T_2=0.7 \mu \mathrm{s}$ describes the fringe damping in the unprotected state. Damping of Ramsey fringes in the protected state ($0<\Delta t<270 \, \mathrm{ns}$) is caused by the $V_g$ pulse jitter rather than dephasing (see the text). }
\end{figure}

Direct measurements of the decoherence time $T_\phi$ in the protected regime, by either Rabi or Ramsey techniques, are not feasible because of vanishing coupling of the qubit to microwave pulses. For this reason we have modified the measurements of Ramsey fringes by analogy with the aforementioned $T_1$ measurements. The pulse sequence is shown in \cref{fig:T2}(a). Both $30 \, \mathrm{ns}$  long $\pi/2$ microwave pulses detuned from the qubit transition frequency by $4 \, \mathrm{MHz}$ are applied in the unprotected state ($q_g=0$), and the qubit is measured after the end of the second pulse. Between the $\pi/2$ pulses, while the qubit underwent free precession, the qubit's protected state is restored by applying a gate voltage pulse ($q_g=1e$). After averaging over 1000 cycles, the Ramsey fringes are recorded by varying the delay between the end of the gate pulse and the second $\pi/2$ pulse. 

Ramsey fringes measured according to this procedure for one of the flux "sweet spots" at $\Phi_{\mathrm{ext}} = 0$ are shown in \cref{fig:T2}(b); the $V_g$ pulse for these measurements is $0.27 \, \mu \mathrm{s}$ long. The difference between the amplitudes of Ramsey fringes at moments $\Delta t =0, \, 0.27 \, \mu \mathrm{s}$ may provide information on dephasing in the protected state if this is the only source of dephasing. However, the accuracy of this technique is limited by the $V_g$ pulse jitter. Indeed, in the rotating frame of the unprotected state, the qubit's state vector rotates in the equatorial plane of the Bloch sphere as soon as the protection is turned on. The angular velocity of these rotations, $\omega=\left(E_{01}^{(0e)}-E_{01}^{(1e)}\right)/\hbar$, is large ($\omega> 2\pi \cdot 1 \mathrm{GHz}$) at both flux sweet spots $\Phi_{\mathrm{ext}} = 0, \, \Phi_0/2$, and even a small jitter can result in a significant error in the position of the qubit's state vector at the end of the $V_g$ pulse. According to the specification, the jitter time of the pulse generator used in our experiments could be as large as $0.3\, \mathrm{ns}$. This jitter-induced phase uncertainty alone, without invoking any dephasing in the protected state, is sufficient to explain the reduced amplitude of Ramsey fringes at $\Delta t=0.28\, \mu \mathrm{s}$. Thus, these measurements can impose only the lower limit on $T_{\phi}$, which is close to $1 \, \mu \mathrm{s}$ for the data in \cref{fig:T2}(b).  Future experiments with better-controlled $V_g$ pulses of different lengths may provide a more detailed information on $T_{\phi}$ at both sweet spots.

It should be noticed that, since the state of the bifluxon qubit is governed by the CPB charge $q_g$, the device is sensitive to the offset charge drifts and quasiparticle poisoning of the CPB island. In order to eliminate the effect of these fluctuations, $q_g$ is measured and, if necessary, re-adjusted to the desired value before each $T_1$ and $T_2$ measurement. For calibration, we tracked one period of the readout dispersive shift oscillations $\delta f_r (V_g)$, with minima and maxima corresponding to integer values of the CPB charge. This measurement allows us to estimate the $q_g$ drift rate to be less than $10^{-2}e/\mathrm{min}$, the quasiparticle tunneling is as rare as 1 event per $30$ min due to the engineered difference between the superconducting gaps in the CPB island and its surroundings (see \cref{sapp:Gap engineering for quasiparticle poisoning mitigation}).

\section{Discussion}
\label{smain:Discussion}

\begin{figure} [t]
\includegraphics[width=\columnwidth]{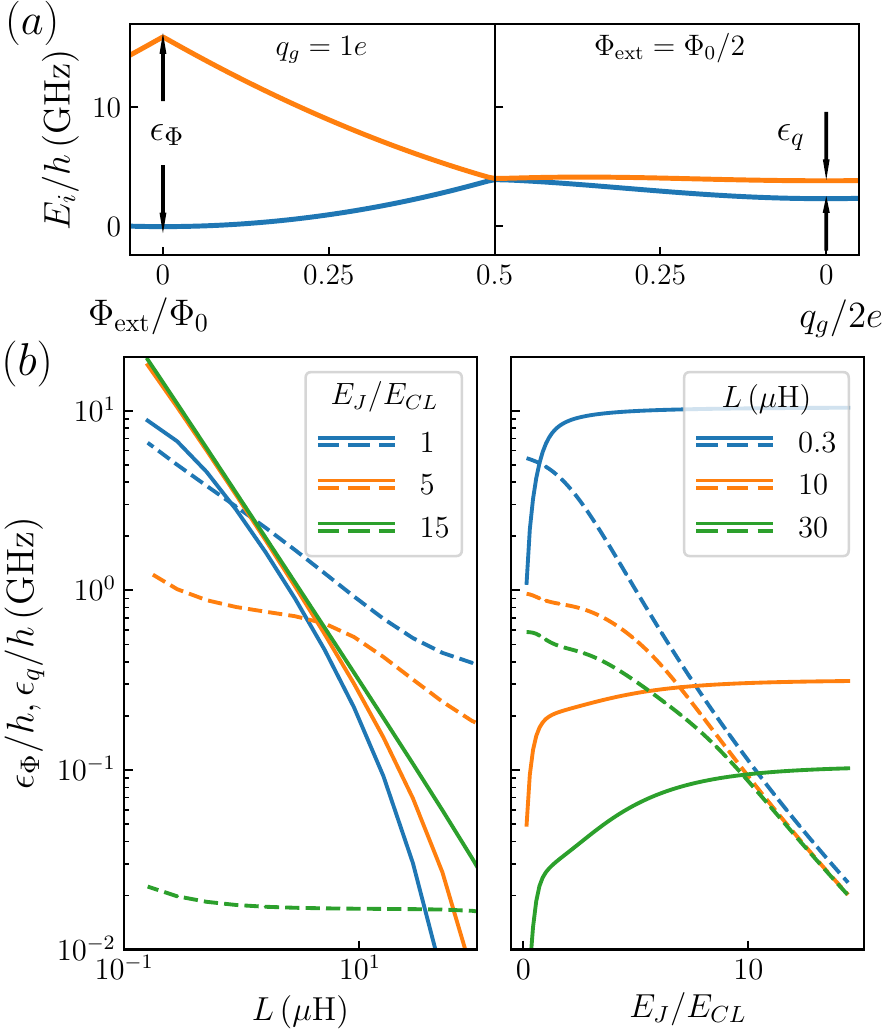}
\caption{\label{fig:theory-2} (a) Two first energy levels of the bifluxon qubit as a function of detuning from degeneracy point. Energy dispersion, which leads to decoherence, can be characterized by amplitudes $\epsilon_{\Phi}$ and $\epsilon_{q}$ [see \cref{eq:dispersion def}, note that $E_{01}(\Phi_0/2, 1e) = 0$ for a symmetric device]. (b) Calculated amplitudes of the flux (solid lines) and charge (dashed lines) energy dispersion as a function of qubit parameters.  }
\end{figure}

\begin{figure*}[t!]
\includegraphics[width=\textwidth]{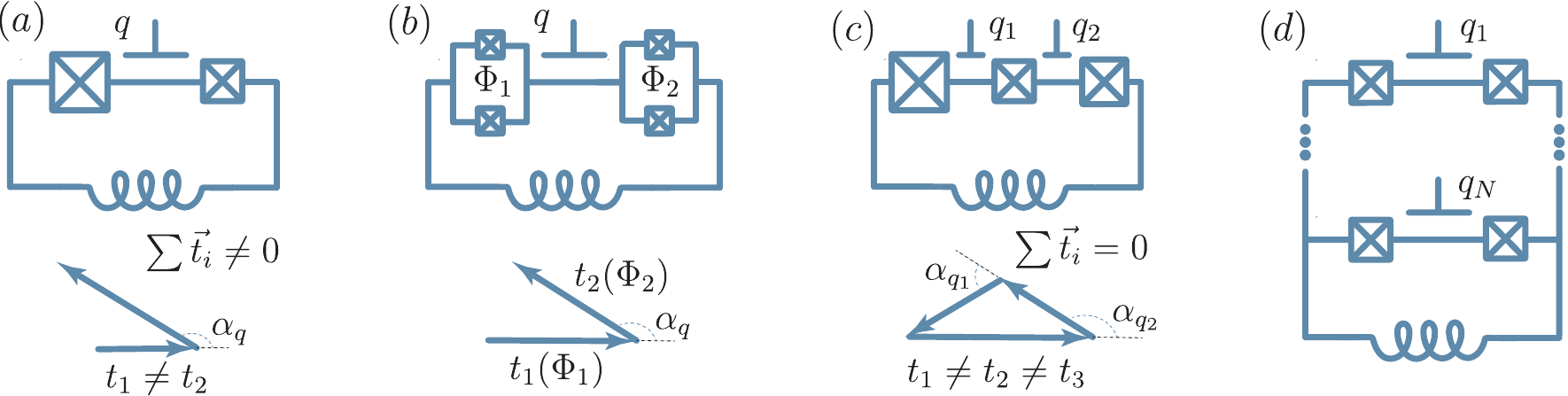}
\caption{\label{fig:bifluxon_imp} Possible ways towards further coherence improvements for the bifluxon qubit. (a) Present design: The asymmetry in the JJs of the circuit leads to distinct complex tunneling amplitudes $t_i$, represented by the vectors at the bottom. As a result, the single phase-slip rate cannot be completely suppressed for any AC phase $\alpha_q = \pi q_g/e $. (b) Device with controllable tunneling probability: disorder of the effective Josephson energies can now be mitigated by the local SQUID frustrations $\Phi_1$ and $\Phi_2$. (c) Adding a second circuit island. The SPS rate can be completely suppressed even for asymmetric junctions as a zero sum of three vectors with comparable lengths. (d) Stacking of the islands into an array: gate charge $q_i = 1e$ on any CPB protects the fluxon parity in the loop, which can be used for expanding the charge sweet spot, similar to Ref. \cite{gladchenko2009superconducting}.  }
\end{figure*} 

In this section we discuss possible modifications to the bifluxon design that could enable further improvement of the qubit coherence beyond the readily available energy decay protection. 


First, let us consider the fully symmetric bifluxon qubit with CPB junctions of identical Josephson energy, where charge noise can still potentially flip the fluxon parity and induce energy relaxation. As we have pointed out earlier, the absolute value of the charge dipole moment [\cref{eq:charge dipole}] is strongly suppressed in comparison to that of a conventional charge qubit. Thus, we find that the condition $E_J/E_{CL} > 10$, similar to the parameter regime of a heavy fluxonium qubit \cite{lin2018demonstration, earnest2018realization}, is in principle enough to achieve $T_1$ times in excess of $10 \, \mathrm{ms}$.

Although the lowest-energy states of the fully symmetric device are exactly degenerate at $\Phi_{\mathrm{ext}}/\Phi_{0} = q_g/2e = 0.5$, deviations from this point open a gap in the spectrum, which leads to decoherence \cref{fig:theory-2}(a). A good measure of the qubit sensitivity to pure-dephasing processes is the amplitude of the charge and flux dispersion of the $0-1$ transition energy $E_{01}(\Phi_{\mathrm{ext}}, q_g)$, defined as

\begin{equation}
\begin{split}
 \epsilon_{\Phi} = E_{01}( 0 , 1e ) &- E_{01}(\Phi_0/2, 1e), \\
 \epsilon_{q} = E_{01}( \Phi_0/2, 0) &- E_{01}(\Phi_0/2, 1e).
\end{split}
\label{eq:dispersion def}
\end{equation}

As it follows from \cref{fig:theory-2}(b), in order to mitigate dephasing due to both charge and flux noises, the optimal strategy is to combine an increase of $E_J/E_{CL}$ with strong reduction of the inductive energy $E_L$. As it was mentioned above, an exponentially small flux dispersion can be achieved in the regime $E_{\mathrm{dps}} \gg 2\pi^2 E_L$. Fulfilling this condition requires the implementation of a ultrahigh-impedance superinductor with $L > 30~ \mu \mathrm{H}$ and self-resonance frequencies $> 1 \, \mathrm{GHz}$. Such an element with a characteristic impedance  $Z > 200 \, \mathrm{k}\Omega$ could be realized by using strongly disordered superconductor nanowires \cite{zhang2019microresonators, rotzinger2016aluminium, coumou2013electrodynamic, niepce2019high, kamenov2019granular} or suspended chains of JJs \cite{pechenezhskiy2019quantum}.

If asymmetry between the CPB junctions is present, the SPS amplitude remains non-zero for any charge on the CPB island [\cref{fig:bifluxon_imp}(a)]. This leads to mixing of the bifluxon states with different parity and increased susceptibility to flux noise. One of the ways to recover the symmetry is to replace the junctions with SQUIDs of a size much smaller than the bifluxon loop area [see \cref{fig:bifluxon_imp}(b)]. This would allow for changing the SQUID's Josephson energy without affecting the optimal flux in the device loop. Alternatively, the SPS can be completely suppressed by introducing a third Josephson junction and an additional gate control line.  Indeed, by independently controlling charges on two CPB islands, the SPS amplitude can be tuned to zero [\cref{fig:bifluxon_imp}(c)].

The sensitivity of a tunable qubit to fluctuations of a control parameter -- the offset charge in our particular case -- is the price to pay for the ability to turn on and off the qubit protection and thus facilitate the gate operations. This sensitivity could be suppressed by combining several qubits in a small array \cite{douccot2012physical}, as it has been demonstrated for the rhombi qubit in Ref. \cite{gladchenko2009superconducting}. In a chain of symmetric Josephson rhombi qubits, the transport of single Cooper pairs is forbidden when $\Phi_{\mathrm{ext}}=\Phi_0 /2$ for any rhombus in the chain. Accordingly, the range of values of $\Phi_{\mathrm{ext}}$ where the qubit is protected (i.e. the size of the sweet spot) increases polynomially with the number of rhombi elements in the chain.
Similarly, a bifluxon qubit made of a small parallel array of CPBs, as shown in \cref{fig:bifluxon_imp}(d), is expected to demonstrate a wider range of $q_g$ tunability for which the  $|g\rangle$ and $|e\rangle$ states remain generate. Realization of such an array would lead to further increase of both the decay and dephasing times beyond the coherence times measured for our proof-of-principle bifluxon-qubit design. 

\section{Conclusion}
\label{smain:Conclusion}

In this work we have developed and characterized a quantum superconducting circuit which serves as a platform for the realization of protected qubits with simultaneous exponential suppression of energy decay from charge and flux noise, and dephasing from flux noise. The circuit is realized as a superconducting loop containing a charge-sensitive Josephson element (a.k.a. Cooper-pair box) and a superinductor. This circuit with two control parameters - the charge on the CPB island and the magnetic flux in the loop - is described by a "two-dimensional" Hamiltonian. Its dimensionality $D>1$ is critical to simultaneous suppression of decay and dephasing via localization of the qubit's wavefunctions in disparate regions of the phase space. The ability to turn the protection on and off by controlling the charge on the CPB island facilitates gate operations with protected qubits. By switching the protection on, we observed a ten-fold increase of the decay time, up to $100\,\mu$s. 
The studied circuit was not expected to demonstrate a long dephasing time because of its sensitivity to fluctuations of charge on the CPB island. However, the bifluxon sensitivity to charge noise is much reduced in comparison with the charge qubit, and the charge-noise-induced dephasing time in the protected state exceeded $1\,\mu$s.
Further improvement of the coherence times can be achieved in the next-generation devices by the optimization of their parameters and combining several $\cos{\left(\phi/2\right)}$ elements in a small array.

\section*{Acknowledgments}

We would like to thank Elio K\"onig, Yashar Komijani, Lev Ioffe and Vladimir Manucharyan for helpful discussions. The work at Rutgers University was supported in part by the NSF awards DMR-1708954, DMR-1838979, and the ARO award W911NF-17-C-0024. The work at the University of Massachusetts Boston was supported in part by a 2019 Google Faculty Research Award and NSF Awards ECCS-1608448, DUE-1723511 and DMR-1838979. The work at the Universit\'e of Sherbrooke was supported in part by the NSERC and the Canada First Research Excellence Fund.

\appendix
\crefalias{section}{appsec}

\section{Derivation of the circuit Hamiltonian}
\label{sapp:Derivation of the circuit Hamiltonian}

\begin{figure} [h!]
\includegraphics[scale=1.0]{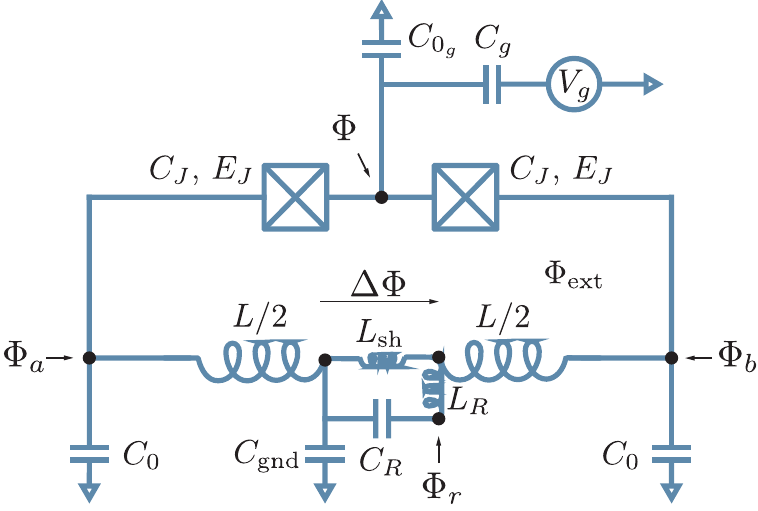}
\caption{\label{fig:bifluxon_device} Schematic diagram of the bifluxon device including the readout resonator, and the gate and stray ground capacitances ($C_0, C_{0_g}, C_{\mathrm{gnd}}$). $V_g$ denotes the gate voltage that controls  the offset charge $n_g^{\varphi}$. $\Delta\Phi/\varphi_0$ is the phase drop across the inductance $L_{\mathrm{sh}}$, which is shared between the qubit and the resonator.}
\end{figure}

In this section, we derive a Hamiltonian for the circuit that includes both the bifluxon qubit and the readout resonator of capacitance $C_R$ and inductance $L_R$ (\cref{fig:bifluxon_device}). Assuming that the phase difference across the inductor shared by the readout resonator and the bifluxon $L_{\mathrm{sh}} \ll L$ is negligible with respect to the total phase drop across the superinductor (see below), we can write an effective circuit Lagrangian of the form
\begin{equation}
\begin{split}
\mathcal{L}&=\frac{C_0}{2}(\dot{\Phi}_a^2 + \dot{\Phi}_b^2) + \frac{C_{0_g}}{2}\dot{\Phi}^2 + \frac{C_g}{2}(\dot{\Phi}-V_g)^2\\
& + \frac{C_J}{2}[(\dot{\Phi}-\dot{\Phi}_a)^2+(\dot{\Phi}-\dot{\Phi}_b)^2]- \frac{1}{2L}(\Phi_a-\Phi_b + \Phi_{\mathrm{ext}})^2\\
& + E_J \Big\{\cos\Big[\frac{2\pi}{\Phi_0}(\Phi-\Phi_a)\Big] + \cos\Big[\frac{2\pi}{\Phi_0}(\Phi-\Phi_b)\Big]\Big\}\\
& + \frac{C_{R}}{2}\dot{\Phi}_r^2 - \frac{1}{2L_R}({\Phi}_r-\Delta\Phi)^2 + \frac{C_{\Delta\Phi}}{2}\Delta\dot{\Phi}^2 - \frac{1}{2L_{\mathrm{sh}}}\Delta\Phi^2,
\end{split}
\label{eq:circuit lagrangian}
\end{equation}
where $\Phi$, $\Phi_a$ and $\Phi_b$ are the circuit flux node variables, $\Phi_r$ and $\Delta\Phi$ are flux branch variables, $\Phi_{\mathrm{ext}}$ is the external flux through the bifluxon-qubit loop, $V_g$ is the gate voltage, and $C_{\Delta\Phi}$ is an effective capacitance for the $\Delta\Phi$ mode. A more convenient basis to treat the qubit Hamiltonian is given by the modes $\Phi_{-} = \Phi_{b}-\Phi_{a}$ and $\Phi_{+} = \Phi_{b}+\Phi_{a}$, in terms of which \cref{eq:circuit lagrangian} reads
\begin{equation}
\begin{split}
\mathcal{L}&=\frac{C_{0_c} + C_g + 2C_J}{2}\dot{\Phi}^2 + \frac{C_0 + C_J}{4}(\dot{\Phi}_{-}^2+\dot{\Phi}_{+}^2) \\
& -C_J\dot{\Phi}\dot{\Phi}_{+}-C_g\dot{\Phi} V_g -\frac{1}{2L}(\Phi_{-}-\Phi_{\mathrm{ext}})^2\\
& + 2 E_J \cos\Big[\frac{2\pi}{\Phi_0}\frac{\Phi_-}{2}\Big]\cos\Big[\frac{2\pi}{\Phi_0}\Big(\Phi-\frac{\Phi_{+}}{2}\Big)\Big]\\
& + \frac{C_{R}}{2}\dot{\Phi}_r^2 - \frac{1}{2L_R}({\Phi}_r-\Delta\Phi)^2 + \frac{C_{\Delta\Phi}}{2}\Delta\dot{\Phi}^2 - \frac{1}{2L_{\mathrm{sh}}}\Delta\Phi^2.
\end{split}
\label{eq:circuit lagrangian v2}
\end{equation}
Since $\Delta\Phi$ is a high-frequency and low-impedance mode, it is assumed to be locked to the semiclassical value $\Delta\Phi\to L_{\mathrm{sh}}\Phi_{-}/(L + L_{\mathrm{sh}})\simeq L_{\mathrm{sh}}\Phi_{-}/L$. Substituting this in \cref{eq:circuit lagrangian v2}, and performing a Legendre transformation, we arrive at the effective circuit Hamiltonian
\begin{equation}
\begin{split}
H&= 4 E_{C_{\varphi}}(n_{\varphi}-n_g^{\varphi})^2 + 4 E_{C_{\phi}}n_{\phi}^2 + 4 E_{C_{\phi_+}}n_{\phi_+}^2\\
& - 2 E_J\cos(\phi/2)\cos(\varphi-\phi_+/2) + \frac{E_L}{2} (\phi - \varphi_{\mathrm{ext}})^2\\\
& + \hbar g_{\varphi\phi_+}n_{\varphi}n_{\phi_+} + \hbar\omega_R a^{\dagger}a + \eta_{\mathrm{sh}}E_L \phi_r\phi 
\end{split}
\label{eq:circuit hamiltonian}
\end{equation}
where we have defined the phase variables $\varphi=\Phi/\varphi_0$, $\phi=\Phi_{-}/\varphi_0$ and $\phi_{+}=\Phi_{+}/\varphi_0$, the respective conjugate charge operators $n_\varphi$, $n_{\phi}$ and $n_{\phi_+}$, and the charging energies $E_{C_{\mu}}=e^2/2C_\mu$ for $\mu\in[\varphi, \phi, \phi_+]$ in terms of the mode capacitances
\begin{equation}
\begin{split}
& C_{\varphi} = C^2/(C_0 + C_J)\\
& C_{\phi} = (C_0 + C_J)/2\\
& C_{\phi_+} = C^2/2(C_{0_g} + C_g + 2 C_J),
\end{split}
\label{eq:effective capacitances}
\end{equation}
with $C^2=(C_{0_g} + C_g)C_J + C_0(C_{0_g} + C_g + 2C_J)$. Note that we neglect a small renormalization of the capacitance of the $\phi$ mode due to $C_{\Delta\Phi}$. We also introduce an effective coupling constant $\hbar g_{\varphi\phi_+}=e^2/2 C_{\varphi\phi_+}$ where $C_{\varphi\phi_+}= C^2/(16 C_J)$, and the offset charge $n^{\varphi}_g = -\beta_\varphi \frac{2eV_g}{8 E_{C_\varphi}}$ with $\beta_{\varphi} = C_g/C_{\varphi}$. The resonator Hamiltonian is written in terms of its resonance frequency $\omega_R$ and the ladder operators ($a, a^{\dagger}$), and we define the inductive participation ratio $\eta_{\mathrm{sh}}=L_{\mathrm{sh}}/L_R$ that quantifies the coupling between the bifluxon qubit and the resonator. 

We assume that $\phi_+$ is a high-frequency mode detuned away from the qubit transitions of interest \cite{hazard2019nanowire}. Under this approximation, the coupling $g_{\varphi\phi_+}$ leads to a small dispersive shift which, however, is not needed to describe the experimental data of the device studied in this work. Under these assumptions, \cref{eq:circuit hamiltonian} reduces to a model of a two-dimensional ($\varphi, \phi$) qubit Hamiltonian coupled to the resonator mode $\phi_r$, which we use in the main text.

Finally, in order to account for the effect of circuit-element disorder on the circuit junctions, we derive a perturbative correction to \cref{eq:circuit hamiltonian} of the form 
\begin{equation}
\delta H = \Delta E_J \sin(\phi/2)\sin(\varphi - \phi_{+}/2)
\label{eq:perturbation by junction disorder}
\end{equation}
where $\Delta E_J = E_{Ja} - E_{Jb}$ is the junction asymmetry, defined in terms of the junction energies $E_{Ja}$ and $E_{Jb}$, and $\bar{E}_J=(E_{Ja}+E_{Jb})/2$. Note that the replacement $E_J\to\bar{E}_J$ in \cref{eq:circuit hamiltonian} should also be made.

\section{Integer charge on the island $n_g = N$}
\label{sapp:Integer charge on the island}

Let us consider the reduced Hamiltonian \cref{eq:reduced H} for the case of an integer charge $n_g$ on the CPB island. In the limit $E_C\gg E_J$, we can restrict the analysis to two charge states. The matrix representation of the Hamiltonian then reads
\begin{equation}
H =  \left( \begin{array}{cc}
            \mathbb{A} + \mathbb{C} & -\mathbb{B}  \\
            -\mathbb{B} & \mathbb{A} - \mathbb{C}  \\
             \end{array} \right),
\label{eq:matrix H}
\end{equation}
where 
\begin{equation}
\begin{split}
\mathbb{A} &= -4 E_{CL}\partial_{\phi}^2 + \frac{E_L}{2}(\phi - \varphi_{\mathrm{ext}})^2, \\
\mathbb{B} &= E_J \cos(\phi /2),\\
\mathbb{C} &= 2 E_{C}. 
\end{split}
\label{eq:params matrix H}
\end{equation}
If $\mathbb{C}$ is the dominant term in \cref{eq:matrix H}, the charge component of the lowest energy eigenvector is close to a pure $|N\rangle$ state
\begin{equation}
|\psi^n\rangle = 
      \left( \begin{array}{c}
             \alpha    \\
             1   \\
             \end{array} \right),
\label{eq:pure N}
\end{equation}
for $\alpha \ll 1$. The eigenvalue
\begin{equation}
\mathbb{E} = \mathbb{A} - \sqrt{ \mathbb{C}^2 + \mathbb{B}^2} \approx \mathbb{A} - \mathbb{C} -\frac{\mathbb{B}^2}{2\mathbb{C}},
\label{eq:eigenvalue N}
\end{equation}
corresponds to a fluxonium-like Hamiltonian of the form
\begin{equation}
H = -4 E_{CL}\partial_{\phi}^2 + \frac{1}{2}E_L(\phi - \varphi_{\mathrm{ext}})^2 - E_J^*\cos\phi,
\label{eq:fluxonium like H}
\end{equation}
where $E_J^* = E_J^2/4E_C$ is a renormalized Josephson energy.
\section{Coupling to the environment and decoherence}
\label{sapp:Coupling to the environment and decoherence}

\begin{figure} [h!]
\includegraphics[scale=1.0]{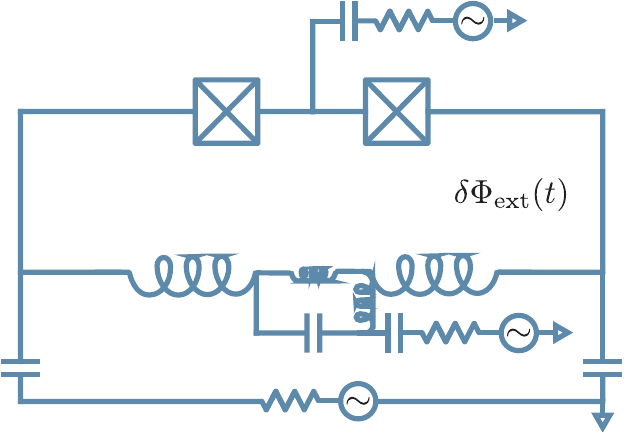}
\caption{\label{fig:bifluxon_environment} Bifluxon device coupled to environmental degrees of freedom leading to decoherence. The resistors model dissipative circuit elements coupled capacitively to the qubit. $\delta\Phi_{\mathrm{ext}}(t)$ represents the magnetic flux fluctuations.}
\end{figure}

In this section, we consider the coupling of the qubit modes to environmental sources of noise, and derive the relaxation rates that are used in the main text to fit $T_1$. \Cref{fig:bifluxon_environment} illustrates the coupling of the bifluxon qubit to external (noisy) degrees of freedom.

Charge-induced decay occurs due to coupling of the Cooper-pair-box variable ($\varphi$) to the environment, mainly via the voltage line that is used to control $n_g^{\varphi}$. The coupling Hamiltonian is of the form $\delta H= 2e n_{\varphi} \beta_{\varphi} \Delta V$, where $\Delta V$ is a voltage-noise operator leading to fluctuations of the offset charge. Using the Fermi's golden rule, we derive the transition rate $\Gamma=|\langle 0|2e n_{\varphi} \beta_{\varphi}|1\rangle|^2 S_{V}(\omega_{01})/\hbar^2\equiv 1/T_1$, where $S_{V}(\omega_{01})$ is the noise spectral density evaluated at the qubit transition frequency. Denoting the impedance of the environment coupled to the qubit port as $Z(\omega)$ and assuming an Ohmic spectral density of the form $S_{V}(\omega) = \hbar\omega \mathrm{Re}[Z(\omega)]\Big[1+\coth\Big(\frac{\hbar \omega}{2 k_B T}\Big)\Big]$ \cite{devoret1995quantum}, we arrive at the expression
\begin{equation}
\frac{1}{T_{1}} = \beta_{\varphi}^2 |\langle 0 |n_{\varphi} | 1\rangle|^2 r_{\mathrm{env}}\,\omega_{01}\Big[1+\coth\Big(\frac{\hbar \omega_{01}}{2 k_B T}\Big)\Big],
\label{eq: T1 CPB}
\end{equation}
where $r_{\mathrm{env}}=\mathrm{Re}[Z(\omega_{01})]/R_K$ is the effective resistance of the electromagnetic environment in units of the reduced superconducting quantum of resistance, $R_K=\hbar/(2e)^2\simeq 1\,\mathrm{k}\Omega$.

Coupling of noise to the fluxonium-like degree of freedom ($\phi$) can be treated similarly. Instead of rewriting \cref{eq: T1 CPB} for $n_{\phi}$, however, we derive an expression that involves the transition matrix elements of the phase operator. This is useful for the discussion of results in the main text. As a consequence of the commutation relation $[\phi,n_{\phi}]=i$, $[\phi,H]=i8E_{C_{\phi}}n_{\phi}$ and thus $\hbar\omega_{01}\langle 0|\phi|1\rangle = i 8E_{C_{\phi}}\langle 0 | n_{\phi}|1\rangle$ \cite{manucharyan2012superinductance}. This relation allows us to rewrite \cref{eq: T1 CPB} as 
\begin{equation}
\frac{1}{T_1} = \beta_{\phi}^2 |\langle 0 |\phi| 1\rangle|^2 r_{\mathrm{env}}\Big(\frac{\hbar\omega_{01}}{8E_{C_{\phi}}}\Big)^2\omega_{01}\Big[1+\coth\Big(\frac{\hbar \omega_{01}}{2 k_B T}\Big)\Big].
\label{eq: T1 fluxonium}
\end{equation}

In order to fit $T_1$, we require the two parameters $\beta_{\varphi,\phi}$ and $r_{\mathrm{env}}$ to be small compared to unity. Moreover, the parameter $r_{\mathrm{env}}$ could in principle have different values in \cref{eq: T1 CPB} and \cref{eq: T1 fluxonium}, because the environment impedance can be different as measured from the multiple qubit ports. 

Finally, we discuss Purcell decay due to coupling of the qubit to the readout resonator. We account for this effect by using a simple model that takes \cref{eq:circuit hamiltonian} into consideration. Rewriting the qubit-resonator coupling as $\delta H=\eta_{\mathrm{sh}}E_L\phi\sqrt{z_r/2}(a+a^{\dagger})$, where $z_r = Z_r/R_K$ is the reduced impedance of the resonator, we follow Ref.~\cite{groszkowski2018coherence} to arrive at
\begin{equation}
\frac{1}{T_{1\mathrm{P}}} = \eta_{\mathrm{sh}}^2\frac{z_r}{2}\frac{\omega_r}{Q_r}\,\frac{(E_L/\hbar)^2}{|\omega_{01}-\omega_r|^2},
\label{eq: T1 Purcell}
\end{equation}
where $Q_r$ is the quality factor of the readout resonator. As expected, the Purcell rate has a significant contribution to $T_1$ only close to the readout resonance frequency (see \cref{fig:T1_summary}). Away from this very narrow frequency range, we find that the qubit relaxation time is very well described by the sum of the two contributions in \cref{eq: T1 CPB} and \cref{eq: T1 fluxonium}.

\begin{figure} [t]
\includegraphics[width=\columnwidth]{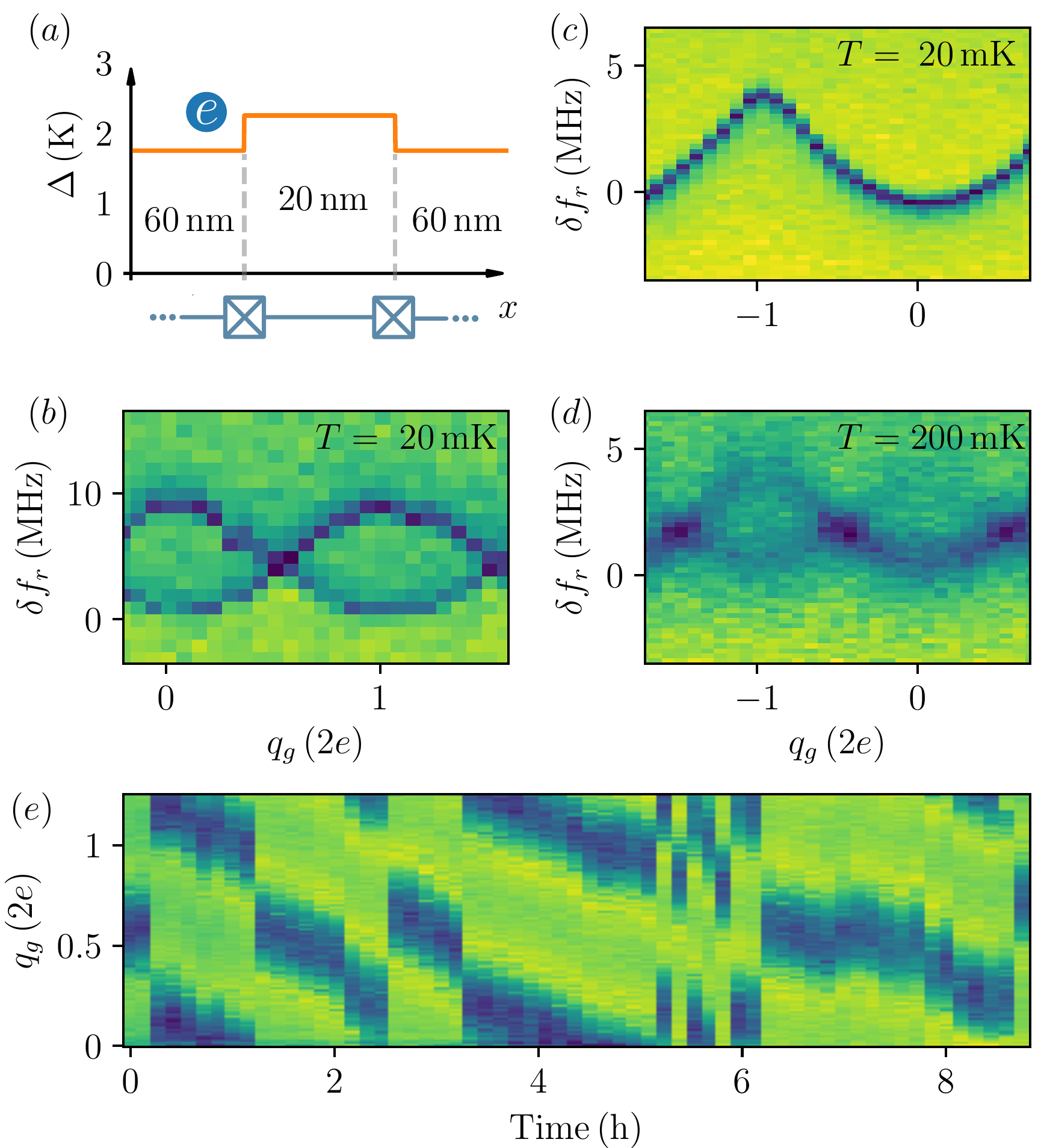}
\caption{\label{fig:qp poisoning} Suppression of quasiparticle poisoning by gap engineering.  (a) Profile of the superconducting gap across the CPB island. The critical temperature of the thin CPB island is by $0.2-0.3 \, \mathrm{K}$ higher than that in the thicker electrodes. (b)-(d) Spectroscopy of the read-out resonator as a function of $q_g$ for bifluxon qubits: without gap modulation at $ 20 \, \mathrm{mK}$ (b), and with gap modulation at (c) $ 20 \, \mathrm{mK}$  and  $ 200 \, \mathrm{mK}$ (d). (e)  The dispersive shift $\delta f_r$ of the readout resonator (color-coded), measured at a fixed gate voltage $V_g$ over $9$ hours. The shift $\delta f_r$ is converted into $\delta q_g$ using the data of panel (c). Abrupt jumps reflect the QP events ($\delta q_g= \pm e$), gradual shift corresponds to a monotonic drift of $q_g$ with the rate $<10^{-2}e/\mathrm{min}$. }
\end{figure}

\section{Gap engineering for mitigation of quasiparticle poisoning }
\label{sapp:Gap engineering for quasiparticle poisoning mitigation}

Quasiparticle poisoning (QP) presents a problem for charge-sensitive quantum superconducting devices \cite{ aumentado2004nonequilibrium, rainis2012majorana}. In particular, for a bifluxon qubit in a protected state, tunneling of a non-equilibrium quasiparticle into/out of the CPB island would remove protection. To minimize QP, we used the so-called gap engineering \cite{ sun2012measurements, ferguson2008quantitative}. \Cref{fig:qp poisoning}(a) shows the superconducting gap in the CPB island and the outer electrodes that form the CPB Josephson junctions. Because of the dependence of the critical temperature of Al films on their thickness, the gap in the thin (20 nm) CPB island is greater than that in thicker (60 nm) outer electrodes. This difference $\delta \Delta$, which we estimate to be $\sim (0.3-0.4) K$, is sufficiently large to block tunneling of non-equilibrium quasiparticles with energies greater than $\delta \Delta$ onto the CPB island at sufficiently low temperatures.

The efficiency of this technique is demonstrated in Figs. 12(b)-(d). If both the CPB island and outer electrodes are thick ($\delta \Delta \simeq 0$), we observe a characteristic "eye" pattern \cite{ sun2012measurements} in the spectroscopic measurements, which reflects rapid $\pm e$ jumps of the CPB charge on the timescale of a single scan of the resonance of the readout resonator, see \cref{fig:qp poisoning}(b). This pattern vanishes if the gap engineering is employed and re-appears only at higher temperatures, where the quasiparticles are thermally excited in the CPB island [compare panels (c) and (d) of \cref{fig:qp poisoning}]. Gap engineering and careful infrared and magnetic shielding of the device allowed us to increase the time intervals between the QP events up to $30 \, \mathrm{mins}$. \Cref{fig:qp poisoning}(e) shows that, in addition to rare QP events, we observed slow monotonic drift of $q_g$ whose origin remains unclear. Because of this drift, we had to measure (and, if necessary, re-adjust) $q_g$ before each time-domain measurement.

\bibliography{bifl}

\end{document}